\documentclass[12pt]{article}
\usepackage{a4}

\usepackage{graphicx}
\newcommand{\fracd}[2]{\frac{\displaystyle #1}{\displaystyle #2}}
\newcommand{\tr}{\bigtriangleup}

\begin{document}

\title{Two-point functions in 4D dynamical triangulation}

\author{Bas V. de Bakker\thanks{email: bas@phys.uva.nl}
  \addtocounter{footnote}{1} \\Jan Smit\thanks{email:
    jsmit@phys.uva.nl} \and Institute for Theoretical Physics,
  University of Amsterdam\\ Valckenierstraat 65, 1018 XE Amsterdam,
  the Netherlands.}

\date{July 6, 1995}

\maketitle

\renewcommand{\abstractname}{\normalsize Abstract}
\begin{abstract}
  \normalsize In the dynamical triangulation model of 4D euclidean
  quantum gravity we measure two-point functions of the scalar curvature 
  as a function of the geodesic distance.  To get the
  correlations it turns out that we need to subtract a squared
  one-point function which, although this seems paradoxical, depends
  on the distance.  At the transition and in the elongated phase we
  observe a power law behaviour, while in the crumpled phase we cannot
  find a simple function to describe it.
\end{abstract}

{%
\thispagestyle{myheadings}
\renewcommand{\thepage}{ITFA-95-1}

\clearpage
}

\section{Introduction}

In the dynamical triangulation model of four dimensional euclidean
quantum gravity the path integral over metrics on a certain manifold
is defined by a weighted sum over all ways to glue four-simplices
together at the faces \cite{AmJu92,AgMi92a}.  This idea was first
formulated in \cite{We82}, using hypercubes instead of simplices.

The partition function of the model at some fixed volume $N$ is
\begin{equation}
  Z(N,\kappa_2) = \sum_{{\cal T}(N_4 = N)} \exp(\kappa_2 N_2).
  \label{partfunc}
\end{equation}
The sum is over all ways to glue $N$ four-simplices together, such
that the resulting complex has some fixed topology which is usually
(as well as in this article) taken to be $S^4$.  The $N_i$ are the
number of $i$-simplices in this complex.  $\kappa_2$ is a coupling
constant which is proportional to the inverse of the bare Newton
constant: $\kappa_2 \propto G_0^{-1}$.

It turns out that the model has two phases.  For low $\kappa_2$ the
system is in a crumpled phase, where the average number of simplices
around a vertex is large and the average distance between two
simplices is small.  In this phase the volume within a distance $r$
appears to increase exponentially with $r$, a behaviour like that of a
space with constant negative curvature.  At high $\kappa_2$ the system
is in an elongated phase and resembles a branched polymer.  As is the
case with a branched polymer, the (large scale) internal fractal
dimension is 2.  The transition between the two phases occurs at a
critical value $\kappa_2^c$ which depends somewhat on $N$.  This
transition appears to be a continuous one, making a continuum limit
possible \cite{AgMi92b,AmJuKr93,CaKoRe94a}.  At the transition, the
space behaves in several respects like the four dimensional sphere
\cite{BaSm95a}.

\section{Curvature and volume}

In the Regge discretization of general relativity, all the simplices
are pieces of flat space.  The curvature is concentrated on the
subsimplices of codimension two, in our case the triangles.  On these
triangles it is proportional to a two-dimensional delta function.
{}From the definition of curvature as the rotation of a parallel
transported vector, one can find the integrated curvature over a small
region $V_\epsilon(\tr)$ around such a triangle
\begin{equation}
  \int_{V_\epsilon(\tr)} R \sqrt{g}\, dx = 2 A_\tr \delta_\tr,
  \label{simpcurv}
\end{equation}
where $A_\tr$ is the area of the triangle and $\delta_\tr$ is the
deficit angle around the triangle (see e.g.\ \cite{Ha86}).  The
deficit angle around a triangle is the angle which is missing from
$2\pi$
\begin{equation}
  \delta_\tr = 2 \pi - \sum_{d \in \{S(\tr)\}} \theta_d
\end{equation}
where $\{S(\tr)\}$ are the simplices around the triangle and
$\theta_d$ is the angle between those two faces of the simplex that
border the triangle.  The angle $\delta_\tr$ can be negative.

In dynamical triangulation, all the simplices have the same size and
shape and the deficit angle is a simple function of the number
$n_{\tr}$ of simplices around the triangle.  Then expression
(\ref{simpcurv}) reduces to
\begin{equation}
  \int_{V_\epsilon(\tr)} R \sqrt{g}\, dx = 2 V_2 (2\pi - \theta
  n_{\tr}),
\end{equation}
where $\theta$ is the angle between two faces of a simplex, which for
$D$ dimensions equals $\arccos(1/D)$, and $V_2 = A_{\tr}$ is the now
constant area of a two simplex.

For each triangle we can define a local four volume that belongs to
the triangle by assigning that part of each adjoining simplex to it
which is closer to the triangle than to any other.  For equal
simplices, this just results in $V_4/10$ per adjoining simplex with
$V_4$ the volume of a four simplex.  In other words this local volume
is
\begin{equation}
  V_\tr = \int_{\Omega(\tr)} \sqrt{g}\, dx = \frac{V_4}{10} n_{\tr},
\end{equation}
where $\Omega(\tr)$ is the region of space associated to that
triangle.  It is not clear what $V_\tr$ would mean in the continuum
limit.  We define it here mainly to compare our results with other
work on simplicial quantum gravity.

If we view the delta function curvature as the average of a constant
curvature over the region $\Omega(\tr)$, this constant curvature would
be equal to
\begin{equation}
  R_{\tr} = \frac{20V_2}{V_4}\, \frac{2 \pi - \theta
    n_{\tr}}{n_{\tr}}.
\end{equation}

Because neither a constant term nor a constant factor is important for
the behaviour of correlation functions, we will in the rest of this
paper for convenience use the definitions
\begin{eqnarray}
R_{\tr} &\equiv & n_{\tr}^{-1},\\
V_\tr &\equiv & n_{\tr}.
\end{eqnarray}

\section{Two-point functions}
\label{twopointsec}

One of the interesting aspects of the dynamical triangulation model
one can investigate is the behaviour of two-point functions of local
observables.  Because we are looking at observables defined on the
triangles we consider correlations between the triangles, at a fixed
distance $d$ which is also defined in terms of triangles.  Such a
correlation function of an observable $O(x)$ will be denoted by
$\langle O O \rangle (d)$.

We define the geodesic distance between two triangles as the smallest
number of steps between neighbouring triangles needed to get from one
to the other.  For this purpose, we define two triangles to be
neighbours if they are subsimplices of the same four-simplex and share
an edge.  Other definitions of neighbour are conceivable. One such a
definition would be to define two triangles to be neighbours if they
share an edge, irrespective of whether they are in the same simplex.
The one we use has the advantage that it is quite narrow and therefore
results in larger distances.

The idea behind our correlation functions is as follows. For each
configuration generated according to the ensemble (\ref{partfunc}), we
take a random pair $(x,y)$ of triangles at distance $d$, where 
$x$ and $y$ denote the triangles. For this pair we calculate the 
observable $O(x)O(y)$.  Then we go to the next configuration and 
repeat the process. Finally we take the average over all such 
pairs. If no such pair exists for a particular configuration, the 
configuration is discarded. 

Obviously, this method would be very inefficient in practice.  We 
improve the statistics by using the average value of $O(x) O(y)$ 
over all pairs $(x,y)$ at distance $d$ in each configuration.
Therefore, we calculate for each configuration
\begin{equation}
  \fracd{\sum_{x,y} O(x) O(y) \delta_{d(x,y),d} }{\sum_{x,y}
    \delta_{d(x,y),d}}.
\end{equation}
Taking the average over configurations
\begin{equation}
  \langle A \rangle = \fracd{\sum_{\cal T} A \exp(\kappa_2 N_2({\cal
      T})) }{\sum_{\cal T} \exp(\kappa_2 N_2({\cal T})) },
\end{equation}
results in the correlation function
\begin{equation}
  \langle O O \rangle (d) = \left\langle \fracd{\sum_{x,y} O(x) O(y)
    \delta_{d(x,y),d} }{\sum_{x,y} \delta_{d(x,y),d} } \right\rangle.
  \label{corrfunc}
\end{equation}

Other definitions are conceivable. One can take random pairs of
triangles from the collection of all configurations.  Configurations
with relatively many pairs of triangles at distance $d$ will then be
counted more often.  In formula, it results in the correlation
function
\begin{equation}
  \langle O O \rangle' (d) = \fracd{\left\langle \sum_{x,y} O(x) O(y)
    \delta_{d(x,y),d} \right\rangle }{\left\langle \sum_{x,y}
    \delta_{d(x,y),d} \right\rangle }.
  \label{altcorr}
\end{equation}
A few experiments did not show a qualitative difference in the
behaviour of (\ref{corrfunc}) and (\ref{altcorr}) at the distances
considered below.  However, for large distances where the finite 
size of the configurations comes into play, the difference 
becomes significant. 

A third possibility, which is natural in Regge calculus, treats 
$n_x$ as a local volume element at $x$ in a discrete 
approximation to a continuum integral over euclidean spacetime:
\begin{equation}
  \langle O O \rangle'' (d) = \fracd{\left\langle 
    \sum_{x,y} n_x n_y O(x) O(y) \delta_{d(x,y),d} 
    \right\rangle }{\left\langle 
    \sum_{x,y} n_x n_y \delta_{d(x,y),d} 
    \right\rangle }.
  \label{altcorr2}
\end{equation}
In this paper we explore the form (\ref{corrfunc}).  We expect that
the correlations constructed from either (\ref{corrfunc}),
(\ref{altcorr}) or (\ref{altcorr2}) will behave identically for not
too large distances.  In a large system, compared to the distance
under consideration, the sum over the triangles will introduce a
self-averaging which probably makes the difference in averaging
between (\ref{corrfunc}) and (\ref{altcorr}) irrelevant.

\begin{figure}[t]
\includegraphics[width=\textwidth]{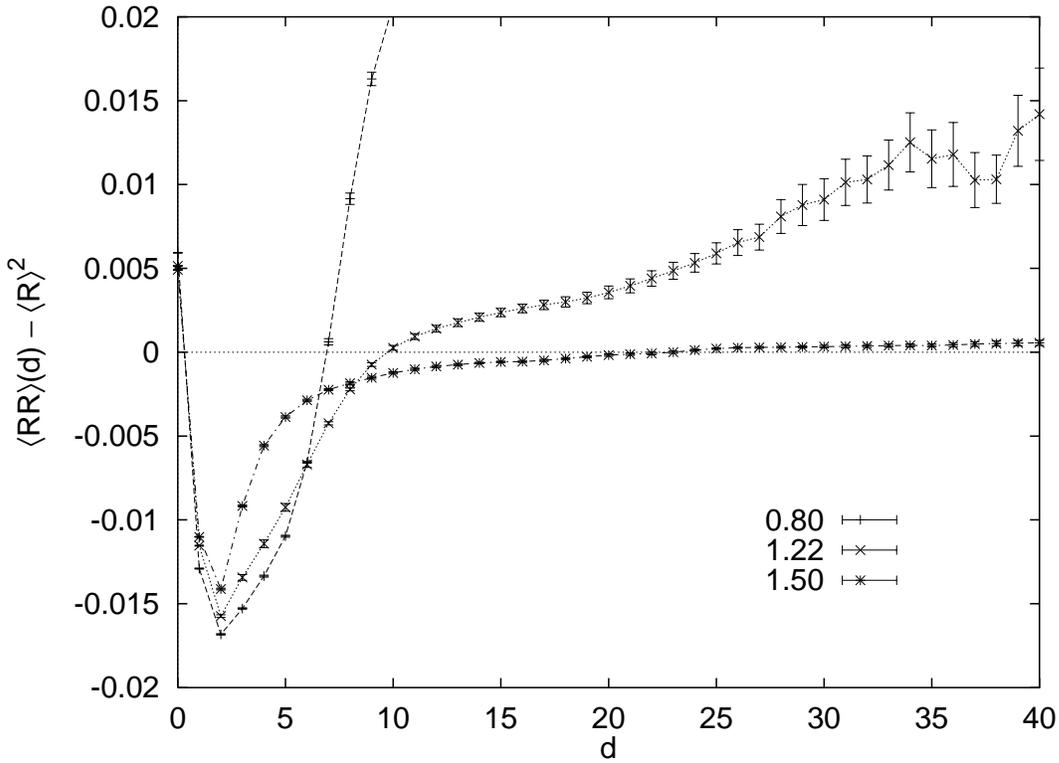}
\caption{The correlation function $\langle R R \rangle (d) -
  \langle R \rangle ^2$ for various values of $\kappa_2$.}
\label{rawrrmafig}
\end{figure}

In figure~\ref{rawrrmafig} we have plotted the correlation function of
the curvature, with the square of its expectation value subtracted.
Most of the data in this paper are for a volume $N = 16000$ simplices.
The values of $\kappa_2$ correspond to a system in the crumpled phase
($\kappa_2 = 0.8$), near (but slightly below) the transition
($\kappa_2 = 1.22$) and in the elongated phase ($\kappa_2 = 1.5$).

Configurations were recorded every $10000$ sweeps, where a sweep is
defined as a number of accepted moves equal to the number of simplices
$N$.  For $\kappa_2 = 0.8$, $1.22$ and $1.5$ we used $16$, $51$ and
$21$ configurations respectively.

One thing is immediately striking, the correlation functions do not go
to zero at long distances.  To keep the short distance behaviour
visible, the full range in the elongated phase has not been plotted,
but we already see that also in this phase it crosses the zero axis
and indeed this curve does eventually go to large ($\approx 0.02$)
positive values.

\begin{figure}[t]
\includegraphics[width=\textwidth]{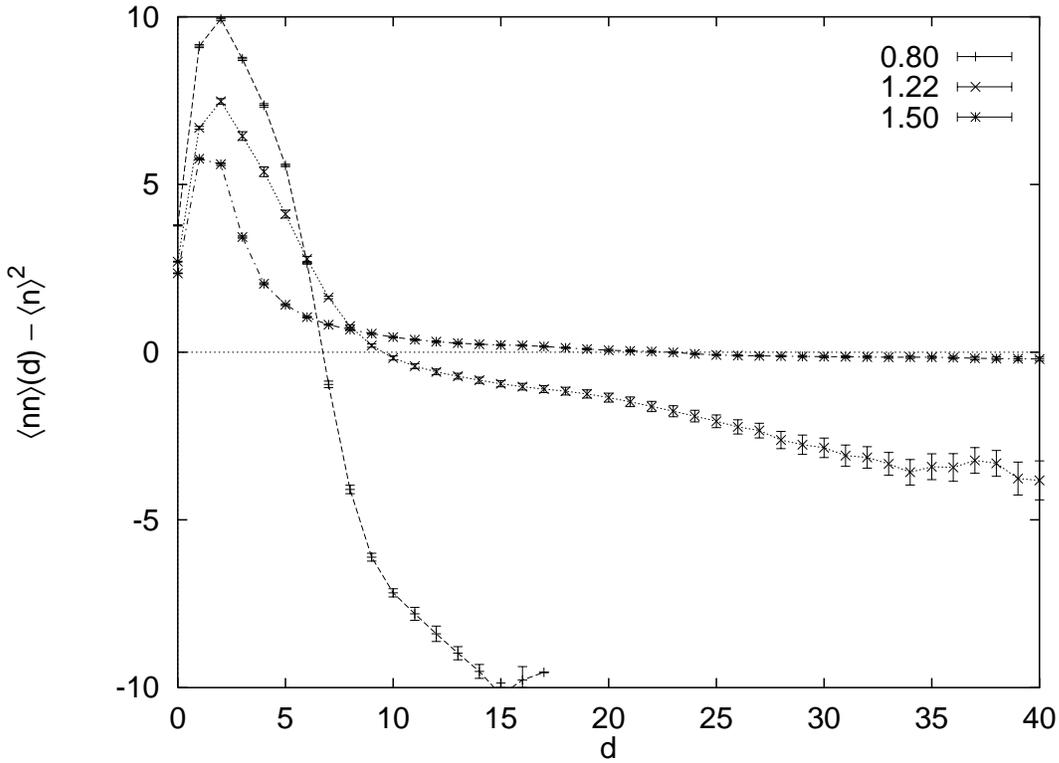}
\caption{The correlation function $\langle n n \rangle (d) -
  \langle n \rangle ^2$ for various values of $\kappa_2$.}
\label{rawggmafig}
\end{figure}

The local volume $V_\tr$ is proportional to the number of simplices
$n_i$ around a triangle $i$.  We see that in this model, this
observable $V_\tr$ is essentially the same as the scalar curvature.
At first sight, one would therefore expect them to have the same
behaviour.  If one is positively correlated, the other one would also
be positively correlated.  Figure~\ref{rawggmafig} shows the
correlation of $n$.  We see that quite the opposite is true.  With few
exceptions, $n$ is positively correlated where $R = n^{-1}$ is
negatively correlated and vice versa.

This behaviour is similar to that reported for the Regge calculus
formulation of simplicial quantum gravity in \cite{Ha94}.  There
it is also found that the curvature correlations are positive and the
volume correlations negative at large distances.

This difference in behaviour can be explained intuitively as follows.
Because triangles with large $n$ have more neighbours, any random
triangle will have a large chance to be close to a point with large
$n$ and a small change to be close to a point with small $n$.  So
whatever the value of $n$ at the origin, the points nearby have large
$n$ and the points far away have small $n$.  The average $ \langle n n
\rangle $ will then be large at small distances and small at large
distances.  Because large $n$ means small $R$, the situation is
reversed if we substitute $R$ for $n$ in this discussion,
qualitatively explaining figs.~\ref{rawrrmafig} and~\ref{rawggmafig}.

At first sight one might conclude from this explanation that a point
with large $n$ having many neighbours is just an artefact of the
model.  This is not true, however.  Large $n$ corresponds to large
negative curvature and also in the continuum a point with large
negative curvature has a larger neighbourhood.  To be more precise,
the volume of $d$-dimensional space within a radius $r$ around a point
with scalar curvature $R$ equals
\begin{equation}
V(R) = C_d r^d (1 - \frac{R}{6(d+2)} r^2 + O(r^4)).
\end{equation}

\section{Connected part}

\begin{figure}[t]
\includegraphics[width=\textwidth]{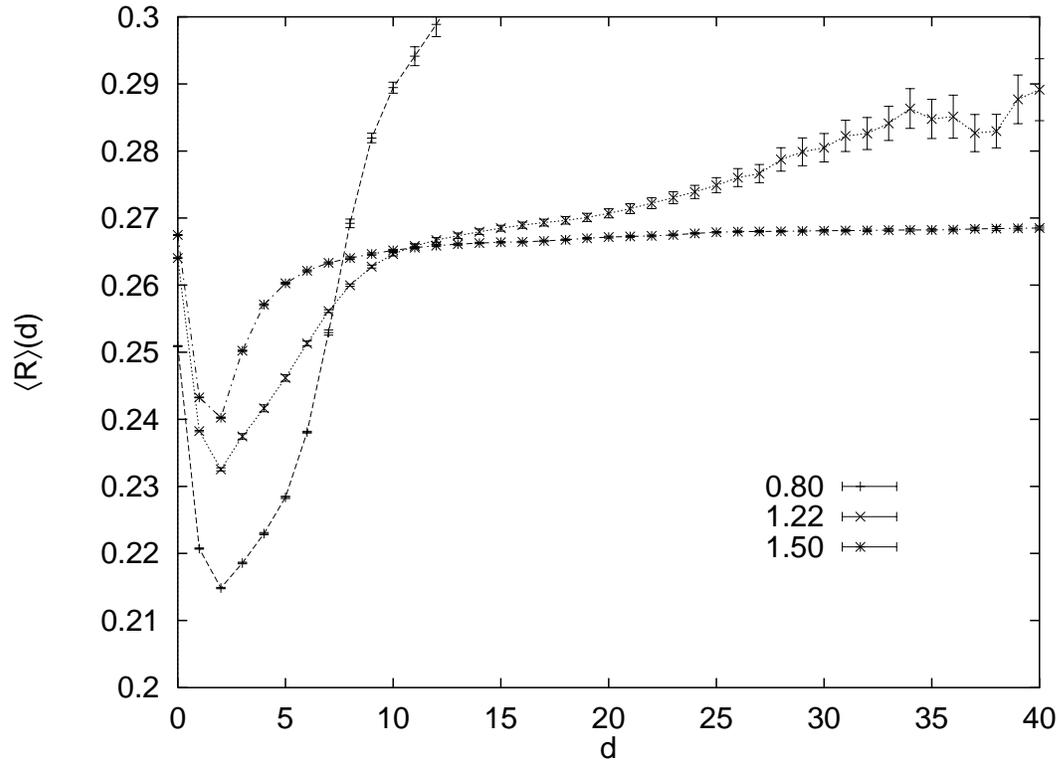}
\caption{The curvature as a function of the distance $\langle R
  \rangle (d)$ for various values of $\kappa_2$.}
\label{rawrfig}
\end{figure}

The above reasoning leads us to the somewhat unusual concept of a
correlation function which does not depend on some observable at the
origin.  We define such a correlation as
\begin{equation}
  \langle R \rangle (d) = \left\langle \fracd{\sum_{x,y} R_x
    \delta_{d(x,y),d} }{\sum_{x,y} \delta_{d(x,y),d} } \right\rangle,
\end{equation}
where $x$ and $y$ denote a triangle.
In the more usual case of a quantum field theory on flat space this
could never depend on the distance, but here it does.  The reason is
that we correlate functions of the geometry with the distance, which
is itself a function of the geometry.

Figure~\ref{rawrfig} shows this correlation function.  No average has
been subtracted.  The behaviour of this one-point function turns out
to be very similar to that of the curvature correlation in
figure~\ref{rawrrmafig}.  This correlation function again shows that
any particular point has a large chance to be in the neighbourhood of
a point with low curvature, which can be simply explained with the
fact that points with low curvature have more neighbourhood.

The same plot for $n$ (not shown) shows the opposite behaviour.  At
small distances it is larger than average, while at large distances it
is smaller than average.  This is rather obvious, because where $n$ is
large, its inverse is small and vice versa.

\begin{figure}[t]
\includegraphics[width=\textwidth]{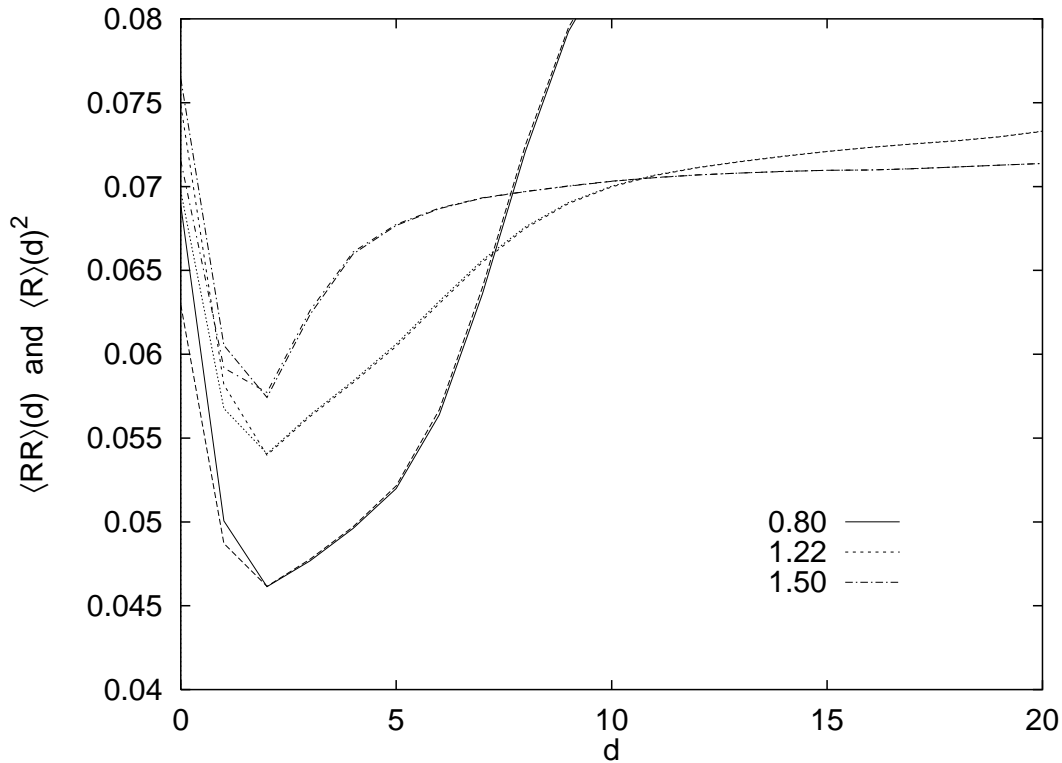}
\caption{Comparison between the correlation function $\langle R R
  \rangle (d)$ (upper curves of each pair) and the squared one-point
  function $\langle R \rangle (d) ^2$ (lower curves) at various values
  of $\kappa_2$.}
\label{rrandr2fig}
\end{figure}

We can now investigate how much of the curvature correlation shown of
figure~\ref{rawrrmafig} is due to this effect.
Figure~\ref{rrandr2fig} compares the curvature correlation $\langle R
R \rangle (d)$ with the square of this one-point function.  We see
that, except at small distances, the two are indistinguishable on this
scale.  In other words, we have not been measuring any curvature
correlations.  All we have measured are correlations between the
curvature and the geodesic distance.  Similarly, $\langle n n \rangle
(d)$ and $\langle n\rangle(d)^2$ are nearly equal.

It is now easy to explain the difference in behaviour between the
curvature and the volume correlations.  Because they are almost equal
to the square of $\langle R \rangle (d)$ and $\langle n \rangle (d)$
respectively, they behave just like them.  And as we just mentioned it
is easy to understand that these have opposite behaviours.

\begin{figure}[t]
\includegraphics[width=\textwidth]{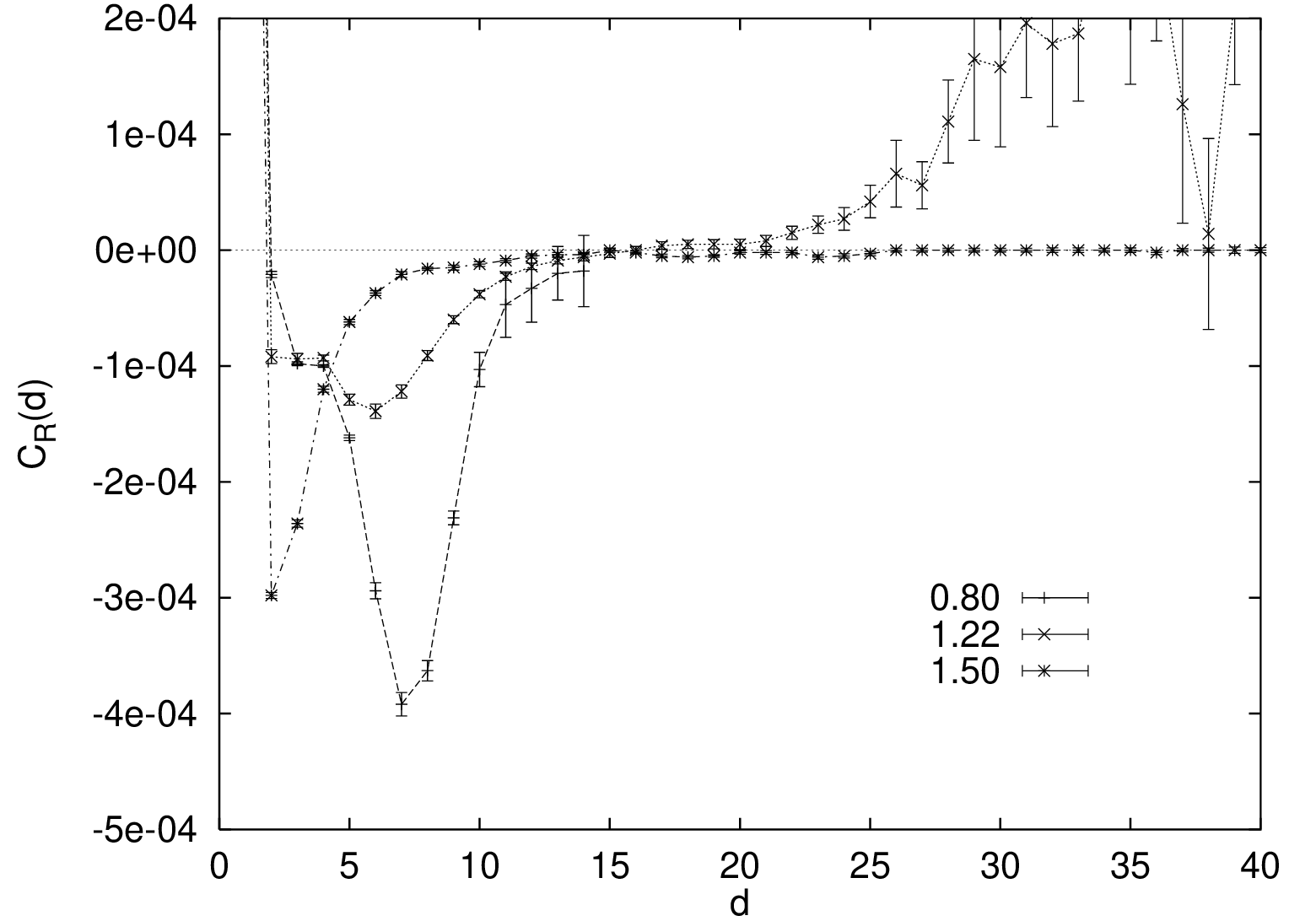}
\caption{Corrected correlation function $C_R(d)$ at various values of
  $\kappa_2$.}
\label{analttfig}
\end{figure}

\begin{figure}[t]
\includegraphics[width=\textwidth]{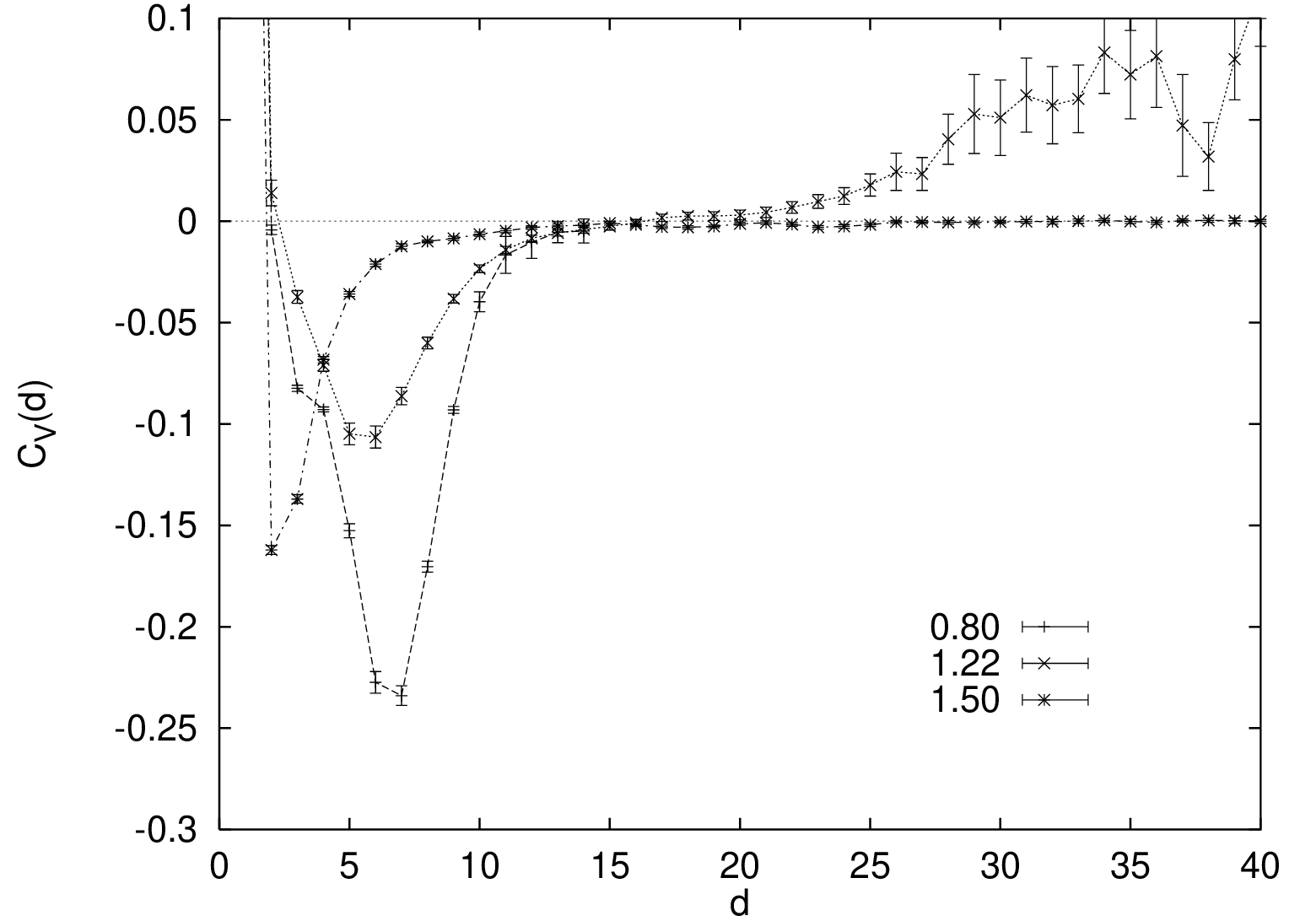}
\caption{Corrected correlation function $C_V(d)$ at various values of
  $\kappa_2$.}
\label{analggfig}
\end{figure}

The way to go now is to subtract the two things and see what real
curvature correlations are left.  This is similar to subtracting a
disconnected diagram and keeping the connected part.  We get the
corrected correlation functions
\begin{eqnarray}
C_R(d) & = & \langle R R \rangle (d) - \langle R \rangle (d) ^2 \\
C_V(d) & = & \langle n n \rangle (d) - \langle n \rangle (d) ^2
\end{eqnarray}
The results for the curvature are plotted in figure~\ref{analttfig}
and those for the volume in figure~\ref{analggfig}.  The errorbars
were found by a jackknife method, each time leaving out one of the
configurations in the calculation of $C_R(d)$ and $C_V(d)$.  Now both
correlations behave almost exactly the same.  Note the large
difference in scale between these figures and figures~\ref{rawrrmafig}
and~\ref{rawggmafig}.

\begin{figure}[t]
\includegraphics[width=\textwidth]{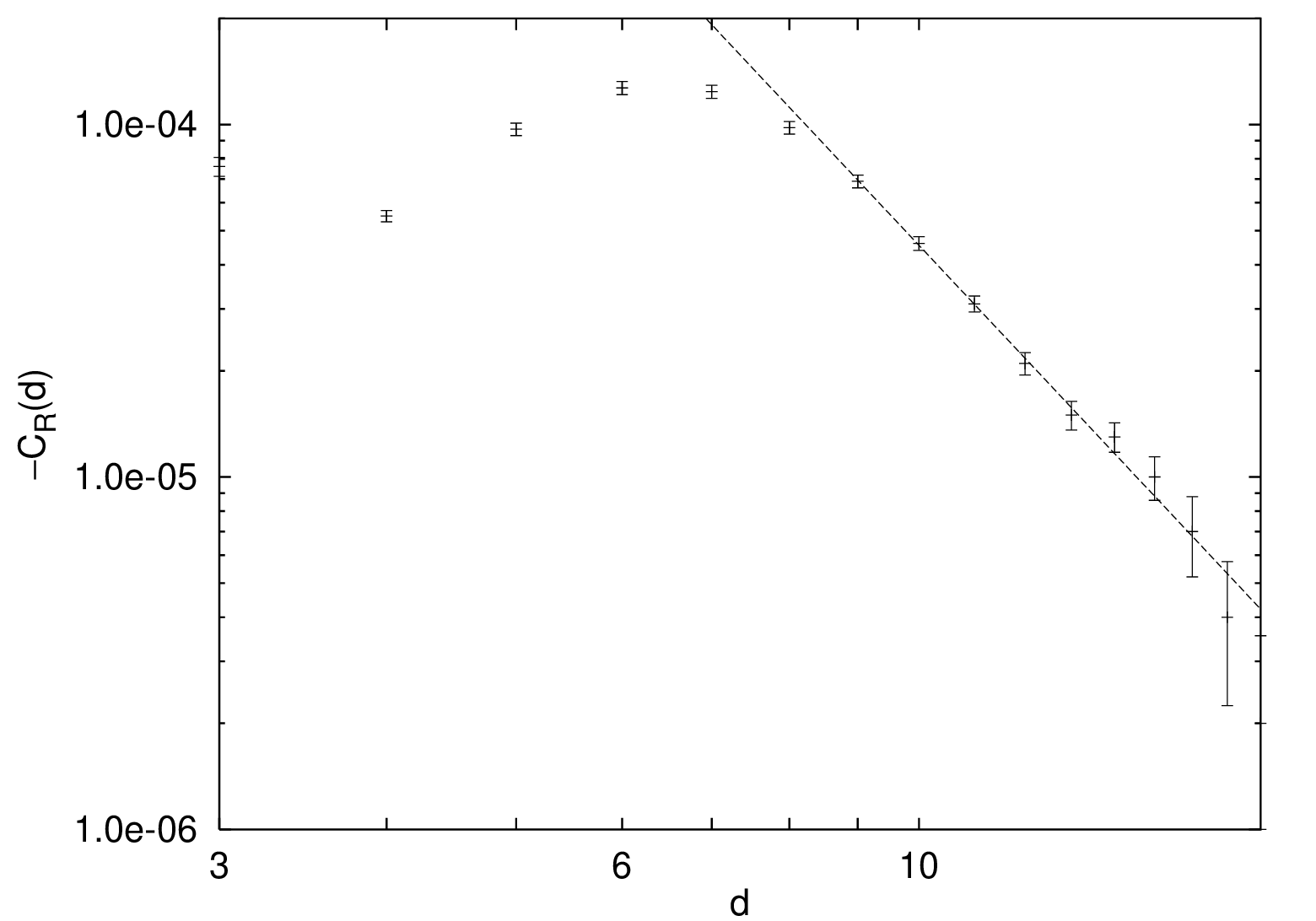}
\caption{Power law fit to curvature correlation $C_R(d)$ near the phase
  transition at $N_4 = 32000$ and $\kappa_2 = 1.255$.}
\label{lltransfig}
\end{figure}

In the crumpled phase we were not able to fit $C_R(d)$ to a simple
function. This is probably due to the fact that we cannot reach very
large distances in this phase.  Near the transition however it is
possible to fit the correlation function to a power law decay, at not
too small distances.  This is shown in figure~\ref{lltransfig}.  In
the region $9 \leq d \leq 18$ it fits nicely to $a d^b$ with the
result
\begin{eqnarray}
  a & = & -0.5(2) \\
  b & = & -4.0(2) \label{transpower} \\
  \chi^2 & = & \mbox{$5$ at $8$ d.o.f.}
\end{eqnarray}
This data was made at a volume of $32000$ simplices, with $\kappa_2 =
1.255$.  We used $65$ configurations, which were recorded every 5000
sweeps.  A similar fit using $C_V$ gives a compatible power,
\begin{eqnarray}
  a & = & -5.7(1.6) \times 10^2 \\
  b & = & -4.30(12) \label{transpower2} \\
  \chi^2 & = & \mbox{$2.3$ at $8$ d.o.f.}
\end{eqnarray}

This result should be taken with caution, however.  One would really
like to have a good fit over a larger range.  To get some idea of the
typical ranges involved, we consider the number of triangles at a
given distance $d$
\begin{equation}
  \langle N'(d) \rangle = \left\langle \fracd{\sum_{x,y}
    \delta_{d(x,y),d}}{N_2} \right\rangle,
\end{equation}
where $N_2$ is the number of triangles of the configuration.  The
corresponding quantity with `triangles' replaced by `four-simplices'
was studied more closely in \cite{BaSm95a}.  The value $d_m$ where
$N'(d)$ has its maximum, is an indication of the distance at which
finite size effects might become important.  At $\kappa_2 = 1.255$,
this $d_m$ is only 11, indicating that finite size effects may play a
role in the measured power.

\begin{figure}[t]
\includegraphics[width=\textwidth]{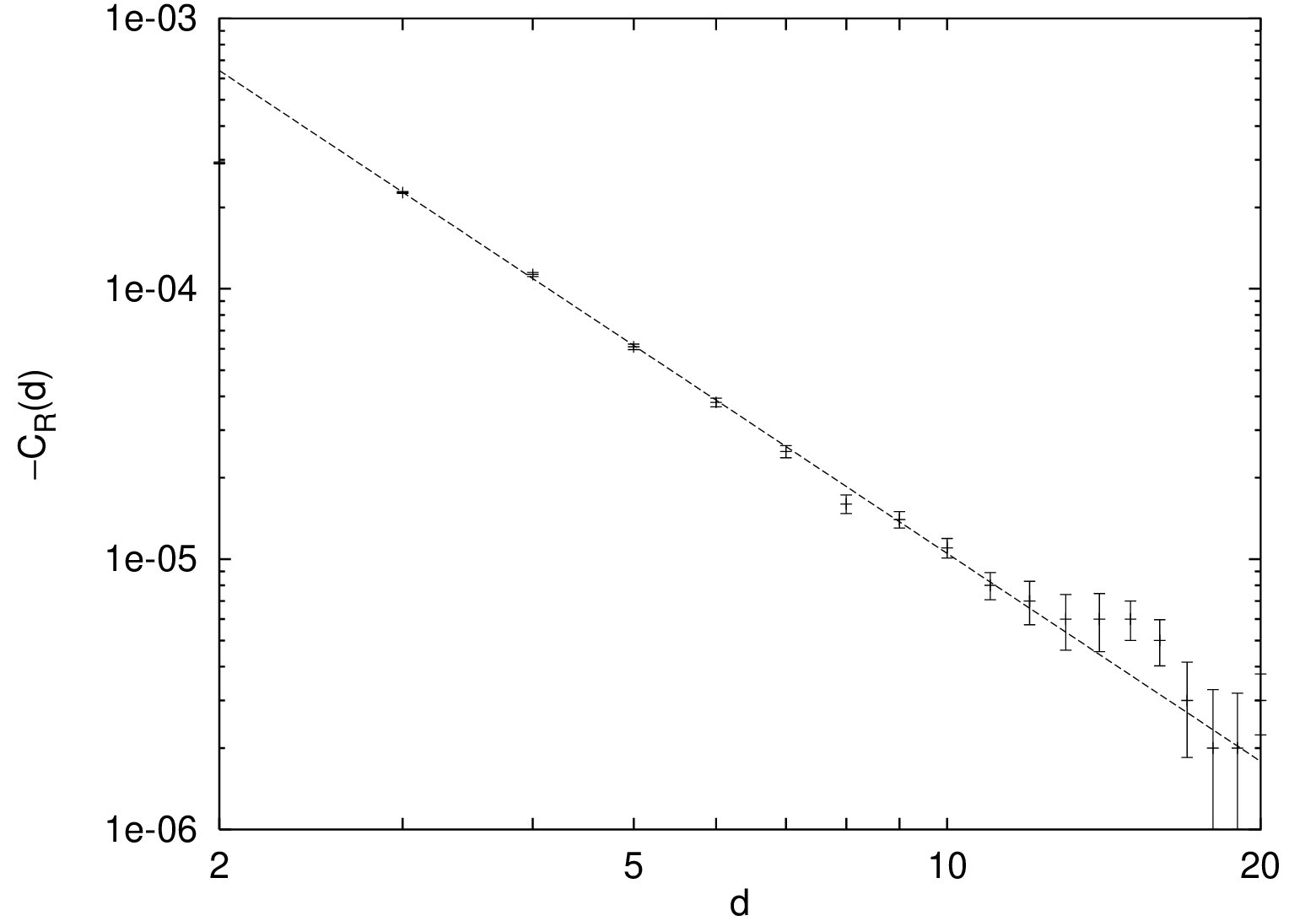}
\caption{Power law fit to curvature correlation $C_R(d)$ in the
  elongated phase at $\kappa_2 = 1.5$.}
\label{llelongfig}
\end{figure}

The situation is even better far in the elongated phase.  Here a power
law fits well, as can be seen in figure~\ref{llelongfig}.  This fit
was done to the points $3 \leq d \leq 15$ and the parameters of this
fit are for $C_R$
\begin{eqnarray}
  a & = & -0.0038(1) \\
  b & = & -2.56(3) \\
  \chi^2 & = & \mbox{$17$ at $11$ d.o.f.}
\end{eqnarray}
and for $C_V$
\begin{eqnarray}
  a & = & -2.2(1) \\
  b & = & -2.57(2) \\
  \chi^2 & = & \mbox{$14$ at $11$ d.o.f.}
\end{eqnarray}

The value of $d$ with maximum number of triangles was 32, so in this
case we are in a region of small distances compared to the system
size.  This data was made from $23$ configurations of $32000$
simplices.  We have also fitted the connected correlation functions at
other points far in the elongated phase and at $16000$ simplices.  The
power that emerged was within the errors equal to the one given above.

\section{Discussion}

We have investigated the behaviour of the curvature and volume
correlation functions.  It turned out that the naive correlation
functions could be almost entirely described by a ``disconnected
part'', which we therefore subtracted.  The difference turns out to
behave according to a power law in the elongated phase and near the
transition.  This indicates the presence of massless excitations.

An obvious question is what the continuum observables are 
corresponding to the correlation functions we have measured.
Generally, in a scaling region, a lattice operator is equivalent 
to a combination of various continuum operators, weighted with 
powers of the lattice distance $a$ according to their dimensions.
In a previous paper \cite{BaSm95a} we found evidence for scaling 
in a surprisingly wide region around the transition value of 
$\kappa_2$. Assuming that the theory can be described by a 
continuum metric tensor $g_{\mu\nu}$ with corresponding curvature 
$R$, the continuum observable with lowest dimension corresponding 
to our lattice correlation function would be given by
\begin{eqnarray}
C_R,\; C_V &\rightarrow &  
    \frac{
    \langle \int dx
    \sqrt{g(x)} \int dy \sqrt{g(y)} \, \delta( d(x,y) - d ) R(x) R(y) 
    \rangle
    }{ \langle
    \int dx \sqrt{g(x)} \int dy \sqrt{g(y)} \, \delta( d(x,y) - d )
    \rangle
    }\nonumber\\
&-& \left(
    \frac{
    \langle \int dx
    \sqrt{g(x)} \int dy \sqrt{g(y)} \, \delta( d(x,y) - d ) R(x)  
    \rangle
    }{ \langle
    \int dx \sqrt{g(x)} \int dy \sqrt{g(y)} \, \delta( d(x,y) - d )
    \rangle
    }
    \right)^2
, \label{continuum1}
\end{eqnarray}
where $d(x,y)$ is the geodesic distance between the points $x$ and $y$
for a given metric $g_{\mu\nu}$. The uniqueness of the lowest dimension 
correlation function in the continuum is in accordance with the 
fact that we found identical behavior for $C_R$ and $C_V$, up to 
an overall factor. Of course, we do not know the effective action 
specifying the average in (\ref{continuum1}). It could be a 
combination of $\int dx \sqrt{g} R$ and higher order $R^2$ terms.

The fact that the connected correlation function is negative 
suggests that it registers fluctuations in the conformal mode of 
$g_{\mu\nu}$.

In a previous paper \cite{BaSm95a} we explored the possibility of a
semiclassical region near the transition, in which the system behaves
like a four-sphere for not too small or large distances.  To this end,
we defined a scale dependent effective curvature.  For $\kappa_2$ near
the transition the following picture emerged.  At small distances,
this effective curvature is large, indicating a Planckian regime.  At
intermediate distances there seems to be a semiclassical regime, where
the space behaves like a four-sphere.  The fluctuations around this
approximate $S^4$ might then correspond to gravitons. We consider it
therefore encouraging that we find the power law behaviour.

For the volumes in current use, the effective curvature shows that the
semiclassical regime sets in at a distance roughly 0.6 of $r_m$
(cf.~fig.~13 in \cite{BaSm95a}).  Here, $r_m$ is the geodesic distance
through the simplices where the number of simplices $N'(r)$ has its
maximum.  We conjectured this fraction to go down at larger volumes.
Similarly, a little beyond 0.6 of $d_m=11$ turns out to be the
distance where the curvature correlations start to behave like
$d^{-4}$ in fig.~\ref{lltransfig}.  We like to think of this as a
confirmation of the point of view sketched above.

Two-point functions of curvature and volume have been studied in the
Regge calculus formulation of simplicial quantum gravity in refs.\ 
\cite{Ha94,BeMaRi94}.  In these studies there are only results in what
is called the well-defined phase of the Regge calculus approach, which
corresponds to our crumpled phase.  This makes it hard to do more than
the qualitative comparison which was done in
section~\ref{twopointsec}.

The curvature correlations have also been investigated in the
continuum. In ref.~\cite{Mod92} they were found to be of zero 
range in the tree approximation to Einstein gravity. To one loop 
order we may expect on dimensional grounds a behavior $G^2 
d^{-8}$ in flat space, and  
$G^2 \overline{R}^2 d^{-4}$ 
for $S^4$, where $G$ is Newton's constant and 
$\overline{R}$ is the $S^4$ background curvature.
    
In \cite{AnMo92} a theory is developed for the conformal factor in
four-dimensional quantum gravity and from this the curvature
correlation is calculated.  The conformally invariant phase discussed
in \cite{AnMo92} seems to correspond to the elongated phase in the
dynamical triangulation model. Intuitively, this can be understood by
visualizing large fluctuations in the conformal factor as generating
many baby universes.  Many baby universes is also a feature of the
branched polymer like elongated phase of simplicial quantum gravity
\cite{AmJaJuKr93}.  Furthermore, the conformally invariant phase would
occur at very large distance scales.  In \cite{BaSm95a} we argued that
the elongated phase also describes scales which are large compared to
a typical physical curvature scale.  In this conformally invariant
phase a power law is predicted for the curvature correlations (see
also \cite{AnMaMo92}).  Unfortunately, a direct comparison with
\cite{AnMo92,AnMaMo92} is not possible because in the continuum the
correlation function is defined as a function of the distance in a
fixed fiducial metric, a quantity that is not yet defined in our
model.  Our result $\approx 2.6$ for the power in the elongated phase
is quite different from the $\approx 0.7$, which corresponds to the
analogue central charge $Q^2 \approx 8$ suggested in \cite{AnMaMo92}.

\section*{Acknowledgements}
The authors would like to thank Emil Mottola for discussions.  We
furthermore thank Piotr Bia{\l}as for useful comments on a first
version of this paper. This work is supported in part by the Stichting
voor Fundamenteel Onderzoek der Materie (FOM).  Most of the numerical
simulations were carried out on the IBM SP1 at SARA.

\end{document}